\documentclass[aps, prd, onecolumn, tightenlines, notitlepage, superscriptaddress, nofootinbib, preprintnumbers, floatfix, showkeys,11pt,a4]{revtex4-2}

\usepackage[normalem]{ulem}
\usepackage{amssymb}
\usepackage{bbold}
\usepackage{amsmath}
\usepackage{graphicx}
\usepackage{url}
\usepackage{ulem}
\usepackage[utf8]{inputenc}
\usepackage{slashed}
\usepackage{lipsum}
\usepackage[T1]{fontenc}
\usepackage{appendix}
\usepackage{csquotes}
\usepackage[x11names,table]{xcolor}

\usepackage{booktabs}
\usepackage{multirow}

\usepackage[export]{adjustbox} 
\usepackage{textgreek} 
\usepackage{caption, subcaption} 
\usepackage{ragged2e} 
\DeclareCaptionJustification{justified}{\justifying}
\usepackage{booktabs} 
\usepackage{orcidlink}
\usepackage{hhline}

\usepackage{hyperref}
\hypersetup{colorlinks,citecolor= nicered,linkcolor= blue}

\definecolor{nicegreen}{rgb}{0.1,0.5,0.1}

\definecolor{myblue}{cmyk}{0.65, 0.37, 0.0, 0.19}
\usepackage{multirow}

\AtBeginDocument{\hypersetup{citecolor=myblue,linkcolor=myblue,urlcolor=myblue}}

\def\be{\begin{equation}}
\def\ee{\end{equation}}
\def\bea{\begin{eqnarray}}
\def\eea{\end{eqnarray}}

\def\cevns{CE\textnu NS}

\def\d{\mathrm{d}}
\newcommand{\qtransfer}{\left|\mathbf{q}\right|}

\newcommand{\AddrAHEP}{%
Instituto de F\'{i}sica Corpuscular (IFIC), CSIC‐Universitat de Val\`encia, E-46980 Valencia, Spain}

\begin{document}
\title{Heavy Sterile Fermions from the Atmosphere: LZ Limits on Neutrino Up-Scattering}%

\author{D. Aristizabal Sierra\orcidlink{0000-0001-5429-3708}}%
\email{daristizabal@uliege.be}%
\affiliation{Universidad T\'ecnica Federico Santa Mar\'{i}a-Departamento de F\'{i}sica\\Casilla 110-V, Avda. Espa\~na 1680, Valpara\'{i}so, Chile}%

\author{V. De Romeri~\orcidlink{0000-0003-3585-7437}}%
\email{deromeri@ific.uv.es}%
\affiliation{\AddrAHEP}%

\author{D. K. Papoulias~\orcidlink{0000-0003-0453-8492}}%
\email{dimitrios.papoulias@uni-hamburg.de}
\affiliation{Institute of Experimental Physics, University of Hamburg, 22761, Hamburg, Germany}

\author{I. M. Shoemaker~\orcidlink{0000-0001-5434-3744}}%
\email{shoemaker@vt.edu}
\affiliation{Center for Neutrino Physics, Department of Physics, Virginia Tech, Blacksburg, VA 24061, USA}%

\begin{abstract}
The LUX-ZEPLIN (LZ) experiment has recently reported data extending to nuclear-recoil energies of about 270~keV, including a single event at 248~keV in a nearly background-free region, for an exposure of 2.84 tonne-years. We examine this dataset through neutrino up-scattering to a heavy sterile fermion induced by atmospheric neutrinos, mediated by vector, axial-vector, scalar and dipole interactions.
Our main result is a set of constraints: using the part of the LZ data where no excess is observed, we obtain new bounds on all four scenarios that probe regions of parameter space not tested before, and show that dark matter detectors can provide competitive limits on neutrino up-scattering for sterile fermion masses above $\sim 50$ MeV. We also test whether the 248~keV event can be explained in this way while remaining consistent with the absence of events at lower recoil energies. For the scalar, vector and axial mediators this requires couplings at the edge of perturbativity, while the dipole does not provide a good fit, so we conclude that up-scattering is not a compelling explanation of the event.
\end{abstract}
\maketitle

\section{Introduction}
\label{sec:intro}
The LUX-Zeplin (LZ) collaboration has recently observed a single nuclear-recoil event at 248$\pm 23$(stat)$\pm 23$(sys) keV in a region of low background, with an exposure of 2.84 tonne-years~\cite{LZ:2026axp}. This has already led to a number of dark matter explanations (see, e.g.,~\cite{Fan:2026kxx,Visinelli:2026kgt,Rodd:2026tyn,McCabe:2026crm,Unwin:2026rdp,Smirnov:2026aqk,deLima:2026shq,Gu:2026vto,Dent:2026bji,Lee:2026wof,Das:2026uyy,Kotlarski:2026pep,Liang:2026coz,DiMauro:2026dqp,Yang:2026wpb,Wang:2026ytg,Bose:2026ndd,Kannike:2026qyl,Bisal:2026khf,Borah:2026zwf,Bandyopadhyay:2026gjw,Du:2026lpa,Ahmed:2026qjg,Alhazmi:2026efz,Okada:2026eol}), as well as a few neutrino-based ones~\cite{Jeesun:2026vzo,Chattaraj:2026fxn,Brdar:2026ukx}. In this paper we focus on the latter, and study what the new LZ data imply for neutrino up-scattering to heavy sterile fermions.

The observation of neutrino masses implies physics beyond the Standard Model (BSM). The most common framework for accounting for them it is to introduce gauge-singlet (``sterile'') fermions. At present, however, we lack sufficient data to determine the mass scale associated with sterile neutrinos, but a wide array of phenomenological probes have been investigated (see~\cite{Abdullahi:2022jlv} for a review). In addition, given that these singlet fermions do not contribute to anomaly cancellation, we lack strong theoretical arguments about the number of them. As such, one should aim to adopt a general framework in the search for sterile fermions. 

There has been significant prior work utilizing neutrino up-scattering in the search for sterile fermions. At low neutrino energies, solar neutrinos at direct detection experiments~\cite{Shoemaker:2018vii,Brdar:2020quo,Shoemaker:2020kji}, and coherent elastic neutrino-nucleus scattering (\cevns) at stopped pion~\cite{Brdar:2018qqj,Candela:2023rvt,Alonso-Gonzalez:2023tgm,Chattaraj:2025rtj,Ahmed:2026knn} and reactors~\cite{DeRomeri:2025csu,Feng:2026jsi,Ahmed:2026knn} provide strong constraints on neutrino upscattering to heavy states. At higher neutrino energies, previous work has used atmospheric~\cite{Gustafson:2022rsz,Atkinson:2021rnp}, and astrophysical neutrino fluxes~\cite{Huang:2022pce}, as well as accelerator neutrino beams at the DUNE near~\cite{Candela:2024ljb,Feng:2026jsi} and far detectors~\cite{Atkinson:2021rnp,Ovchynnikov:2022rqj}. Collider signal predictions of such states have been as well considered, see e.g. Ref. \cite{Bondarenko:2018ptm}. Although typically relying on active-sterile mixing, these analyses and the corresponding searches cover regions in parameter space not accessible to neutrino scattering experiments. Since a 248~keV nuclear recoil can only be induced by the high-energy tail of the atmospheric flux, the LZ event probes up-scattering at higher energies than the solar, reactor and stopped-pion searches, and to heavier sterile masses.

In what follows, rather than restricting the neutrino-sterile-fermion interaction to a single operator, we study four representative simplified models for up-scattering on nuclei, with a vector, scalar, axial or dipole neutrino-sterile-fermion bilinear. In the first three, the interaction is mediated by a new state with generic couplings to the nucleons, whereas the dipole is a transition magnetic moment coupling to the photon. Our main goal is to determine what the new LZ data tell us about these scenarios. Using the low-energy part of the spectrum, where LZ observes no excess over the background expectation, we obtain constraints on each scenario that probe new regions of parameter space, especially for heavy sterile fermions. We then ask, more tentatively, whether up-scattering of atmospheric neutrinos can also account for the one-event observation without conflicting with these constraints, and we infer the preferred regions of parameter space.

The remainder of this paper is organized as follows. In Sec.~\ref{sec:theoretical_frame} we describe the scenarios considered, together with the formulas required for our calculations. In Sec.~\ref{sec:results} we present our results: first, the regions of parameter space ruled out by the LZ data in the first bin, where no excess over the background is observed; and second, the regions in which up-scattering of atmospheric neutrinos could account for the one-event observation, if any. Finally, in Sec.~\ref{sec:conclusions} we summarize and present our conclusions.
\medskip

\section{Theoretical framework}
\label{sec:theoretical_frame}
We consider the up-scattering $\nu\,\mathcal{N}\to\chi\,\mathcal{N}$ of an active neutrino into a heavy gauge-singlet (sterile) fermion $\chi$ of mass $m_\chi$, on a nucleus $\mathcal{N}$ of mass $m_{\mathcal{N}}$. This process has a kinematic threshold, thus only the highest energies will be able to produce scattering at all.  This of course does not preclude solar neutrinos to do so, however the recoil energies recorded by LZ, up to a few hundred keV, can only be reached by the high-energy tail of the atmospheric flux. In xenon, $^8$\text{B} and $hep$ neutrinos can at most induce recoil energies of the order of 5 keV. 

A generic construction of up-scattering processes can involve any possible Lorentz-invariant fermion bilinear. Here, we  consider vector, axial and scalar up-scattering ($a =$ V, A, S) off the xenon nucleus. Their interactions are dictated by the following set of operators
\begin{equation}
    \label{eq:LI_operators}
    \mathcal{L}=g_a^\nu\overline{\nu}\Gamma_a\chi X^a
    + g_a^q\overline{q}\Gamma_a q X^a + \text{H.c.}\ ,
\end{equation}
where $\Gamma_a=\{\gamma^\mu,\gamma^\mu\gamma_5,\mathbb{1}\}$, and $X^a\subset \{V^\mu,\mathrm{A}^\mu,S\}$ is the corresponding vector, axial-vector or scalar mediator. We assume universal couplings to all quark and neutrino flavors, $g_a^\nu=g_a^q\equiv g_a$ (quark-lepton universality). Note that in the limit $\qtransfer^2\ll m_X^2$ these interactions reduce to neutrino generalized interactions (NGI)~\cite{AristizabalSierra:2018eqm}. Although these couplings should arise from full theoretical models, rather than specifying their origin, the Lagrangian is used as a low-energy phenomenological description.

With these assumptions, the differential vector- and scalar-mediated cross sections in terms of the nuclear recoil energy $T_\text{nr}$ are respectively \cite{Candela:2023rvt,Candela:2024ljb}
\be
    \label{eq:vector_xsec}
    \frac{\d \sigma^\text{V}}{\d T_\text{nr}}  = \frac{m_{\mathcal{N}}C_{V}^{4}}{2\pi (m_{V}^{2} + \qtransfer^2)^2}
    \left[ \left( 1-\frac{m_{\mathcal{N}}T_\text{nr}}{2E_{\nu}^{2}}- \frac{T_\text{nr}}{E_{\nu}}+ \frac{T_\text{nr}^{2}}{2E_{\nu}^{2}}\right)-\frac{m_{\chi}^{2}}{4E_{\nu}^{2}}\left(1+\frac{2E_{\nu}}{m_{\mathcal{N}}}-\frac{T_\text{nr}}{m_{\mathcal{N}}}  \right) \right] \ ,
\ee
and 
\be
    \label{eq:scalar_xsec}
    \frac{\d \sigma^\text{S}}{\d T_\text{nr}}  = \frac{m_{\mathcal{N}}C_{S}^{4}}{4\pi (m_{S}^{2} + \qtransfer^{2})^2}\left( 1+ \frac{T_\text{nr}}{2m_{\mathcal{N}}}\right)\left(\frac{m_{\mathcal{N}}T_\text{nr}}{E_{\nu}^{2}} + \frac{m_{\chi}^{2}}{2E_{\nu}^{2}}\right)\, ,
\ee
where $\qtransfer = \sqrt{2m_{\mathcal{N}}T_\text{nr}}$ denotes the magnitude of the three-momentum transfer, $E_\nu$ is the incoming neutrino energy, and $m_{V,S}$ are the mediator masses. In the vector case, and because of the quark-lepton universality assumption, the vector coupling has a rather simple form
\begin{equation}
    \label{eq:CV}
    C_{V}^{2} \equiv 3  g_{V}^{2} \left(Z F_p^M (\qtransfer^2) + N F_n^M(\qtransfer^2) \right)\ ,
\end{equation}
where $Z$ and $N$ are the nuclues proton and neutron numbers, respectively, and the nuclear form factors $F_{p,n} (\qtransfer^2)$ are obtained via the \emph{M-response} (Coulomb operator) in shell model calculations performed in Refs.~\cite{Hoferichter:2018acd,Hoferichter:2020osn}. The scalar coupling is instead isospin-violating, since it is set by the quark content of the nucleon mass,
\be
    \label{scalar_coupling}
    C_{S}^{2} \equiv g_{S}^{2} \left(Z F_p^M (\qtransfer^2)\sum_{q =u,d}\frac{m_{p}}{m_{q}}f_{T_{q}}^{(p)} + N F_n^M(\qtransfer^2)\sum_{q =u,d}\frac{m_{n}}{m_{q}}f_{T_{q}}^{(n)} \right)\ ,
\ee
$m_{p,n}$ being the proton or neutron mass, and $m_{q}$ the quark mass; we neglect the contributions of the strange quark and of the heavy quarks $c,b,t$ (the latter entering through the gluon content of the nucleon).
The couplings $f_{T_q}^{p,n}$ are derived using chiral perturbation theory, and their theoretical uncertainties are within the 10\% level, see e.g. Ref. \cite{Cheng:2012qr}. For our calculation we employ values from Ref. \cite{DelNobile:2021wmp},
\begin{align}
    \label{eq:scalar_fTq}
    f_{T_u}^p&=0.026\ ,\qquad f_{T_d}^p=0.038\ ,
    \nonumber\\
    f_{T_u}^n&=0.018\ ,\qquad f_{T_d}^n=0.056\, . 
\end{align}

Regarding the nuclear form factors, it is worth noting that they can also be approximated with phenomenological approximations. A commonly adopted choice is the Klein-Nystrand parametrization given by \cite{Klein:1999qj}
\begin{equation}
    \label{eq:KN_param}
    F_\mathrm{KN}(\qtransfer^2)=\frac{3 j_1(\qtransfer R_A)}{\qtransfer R_A(1 + a_k^2\qtransfer^2)}\ ,
\end{equation}
where $j_1(x)$ is the first-order spherical Bessel function of the first kind and for the Yukawa potential range we adopt $a_k=0.7\,$fm. In this case, for the nuclear-charge radius we use the parametrization $R_A=1.23\,A^{1/3}\,\text{fm}$ (with $A$ the target mass number). Either choice, the nuclear shell-model or the KN parametrization, carries an intrinsic theoretical uncertainty~\cite{AristizabalSierra:2019zmy,Papoulias:2019lfi,Sierra:2023pnf}.

\begin{figure}[t]
    \centering
    \includegraphics[width=\linewidth]{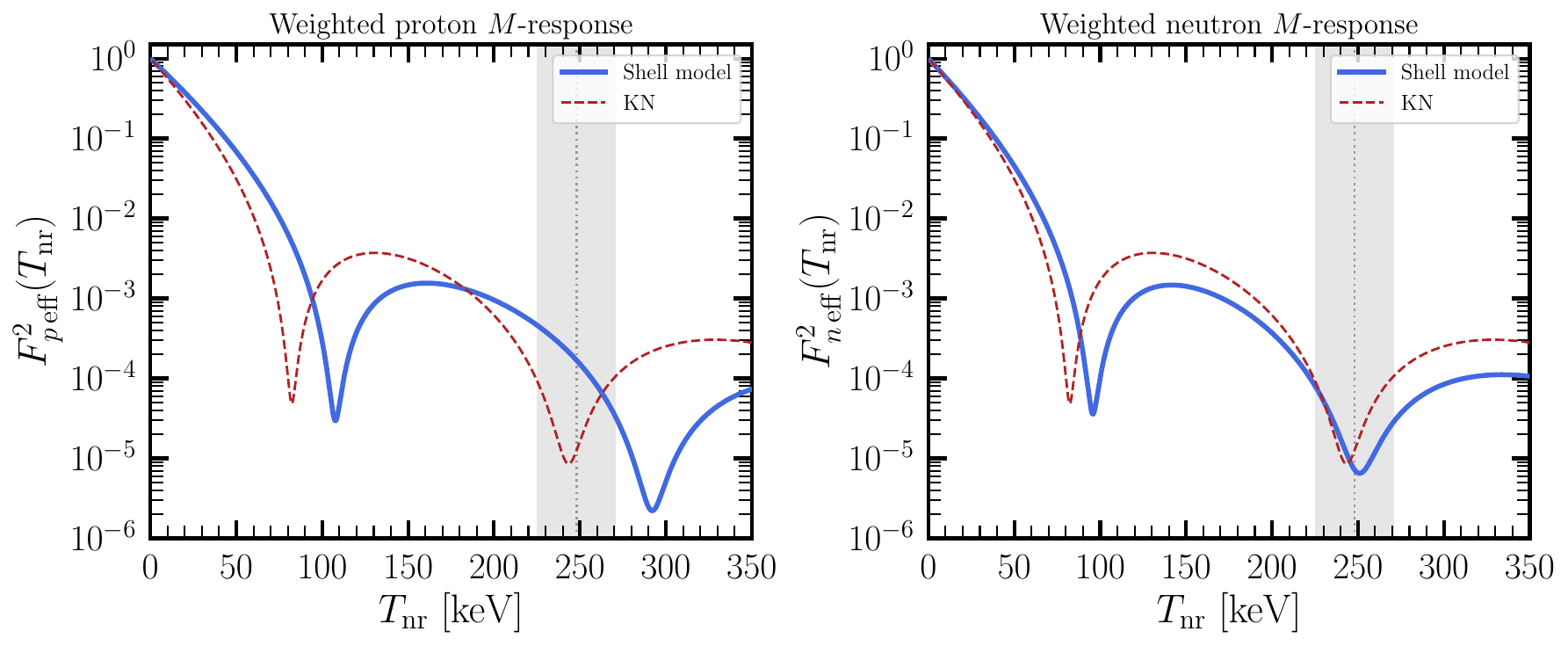}
   \caption{Abundance-weighted effective proton (left) and neutron (right) form factors squared of natural xenon, $F_{p,\mathrm{eff}}^2$ and $F_{n,\mathrm{eff}}^2$ of Eq.~\eqref{eq:effective_xe_ff}, as a function of the nuclear recoil energy $T_\text{nr}$, for the Klein--Nystrand (KN) parametrization and the shell-model calculation. The nine stable isotopes $^{124,126,128,129,130,131,132,134,136}$Xe are combined according to their natural abundances. The dotted vertical line marks $T_\text{nr}=248$~keV, and the shaded band indicates the signal window $225$--$270$~keV used in the analysis.}
    \label{fig:form_factor}
\end{figure}

Phenomenological form factors generally place their diffraction minima at different momentum transfers than nuclear-structure calculations, although they can be tuned to resemble the shell-model prediction more closely. This is relevant here because the 248~keV LZ event lies close to the second minimum of the nuclear form factors: for $A=131$ the KN form factor vanishes at $T_\text{nr}\simeq 244$~keV, but for natural xenon it spans roughly across the range $230$--$270$~keV. For this reason, in this work we compare\footnote{Ref.~\cite{Hoferichter:2018acd} provides no shell-model form factors for $^{124,126}$Xe, for which we use KN instead; together they account for only 0.1842\% of natural xenon.} KN and shell-model form factors~\cite{Hoferichter:2018acd}, weighting all stable isotopes $\mathrm{^{124,126,128,129,130,131,132,134,136}Xe}$ by their natural abundances $\eta_i$.  To avoid one plot per isotope, Fig.~\ref{fig:form_factor} shows the abundance-weighted effective proton and neutron form factors
\begin{equation}
\label{eq:effective_xe_ff}
F_{p,\mathrm{eff}}^2(T_{\rm nr})
=\sum_i\eta_i F_{p,i}^2(\qtransfer_i^2),
\qquad
F_{n,\mathrm{eff}}^2(T_{\rm nr})
=\sum_i\eta_i F_{n,i}^2(\qtransfer_i^2)\,,
\end{equation}
where $\qtransfer_i^2=2m_{\mathcal N_i}T_{\rm nr}$.
Interestingly, the second diffraction minima of the KN and shell model form factors occur at different recoil energies, particularly for the proton response, which is especially relevant for the dipole portal discussed below.

The axial differential cross section requires a more detailed discussion. In contrast to vector and scalar interactions---which are intrinsically coherent---axial interactions are spin-dependent, and require a considerably more involved nuclear response. For the case of natural Xe, only $^{129} {\rm Xe}$ (with spin $J=1/2$) and $^{131} {\rm Xe}$ ($J=3/2$) contribute to axial scattering, with relative abundances of 26.4\% and 21.3\%, respectively. The differential cross section for a neutrino scattering on a nucleus with nuclear ground state $J$ is given by
\begin{equation}
\label{eq:axial_xsec}
\begin{aligned}
\frac{\d\sigma^\text{A}}{\d T_\text{nr}}
&= \frac{2m_\mathcal{N}}{2J+1}
\frac{g_A^4}{(m_\mathrm{A}^2+2m_\mathcal{N} T_\text{nr})^2}
\Bigg\{
\left[
\left(
2+\frac{m_\mathcal{N} T_\text{nr}}{E_\nu^2}
-\frac{2 T_\text{nr}}{E_\nu}
\right)
-\frac{m_\chi^2}{2E_\nu^2}
\left(
1+\frac{3E_\nu}{m_\mathcal{N}}
+\frac{m_\chi^2}{m_\mathcal{N} T_\text{nr}}
\right)
\right] S^\mathcal{T}(\qtransfer^2)
\\[4pt]
&\qquad+
\frac{m_\chi^2}{E_\nu^2}
\left(
2+\frac{3E_\nu}{m_\mathcal{N}}
+\frac{m_\chi^2}{m_\mathcal{N} T_\text{nr}}
\right) S^\mathcal{L}(\qtransfer^2)
\Bigg\}\, ,
\end{aligned}
\end{equation}
where $m_\mathrm{A}$ is the axial-mediator mass, $g_\mathrm{A}$ is the coupling appearing in Eq.~\eqref{eq:LI_operators}. The two structure functions,  $S^\mathcal{T}$ and $S^\mathcal{L}$, involve transverse (\emph{$\Sigma'$-response}) and longitudinal (\emph{$\Sigma''$-response}) multipoles. In contrast to the axial component of neutrino-nucleus scattering, the longitudinal structure function also contributes, with a kinematic coefficient proportional\footnote{Note that, although not displayed in Eq.~\eqref{eq:axial_xsec}, the full longitudinal contribution contains additional terms independent of \(m_\chi\), which remain nonzero even for massless final states. These contributions are suppressed by \(T_{\rm nr}/m_{\mathcal N}\), \(T_{\rm nr}^2/E_\nu^2\), and higher-order recoil corrections.} to $m_\chi^2$.  Interestingly, in the heavy-sterile regime, the transverse kinematic coefficient approximately vanishes at $E_\nu\simeq\sqrt{2}E_\nu^{\min}$ [see Eq.~\eqref{eq:Evmin_kinem}]. Consequently, for a fixed upper neutrino-energy cutoff, this cancellation is accessible only for $m_\chi\lesssim\sqrt{\sqrt{2}\qtransfer E_\nu^{\max}}$, i.e., a slightly more restrictive condition than the kinematic production threshold (see discussion below).

The structure functions are combinations of proton and neutron isoscalar and isovector responses, associated with the longitudinal and transverse-electric operators together with the transverse-magnetic one (see e.g.~\cite{Chasioti:2009fby,Hoferichter:2020osn}). Explicitly,
\begin{equation}
    \label{eq:structure_functions}
    S^{\mathcal{L}, \mathcal{T}}=
    a_0^2\, S_{00}^{\mathcal{L}, \mathcal{T}}
    + a_0 a_1\, S_{01}^{\mathcal{L}, \mathcal{T}}
    + a_1^2\, S_{11}^{\mathcal{L}, \mathcal{T}}\ ,
\end{equation}
where the $S_{ij}^{\mathcal{L}, \mathcal{T}}$ are obtained from nuclear shell-model calculations (see Refs.~\cite{DelNobile:2021wmp,Hoferichter:2020osn}). The isoscalar and isovector nucleon combinations $a_0$ and $a_1$ are determined by the quark spin contents $\Delta_q^{(N)}$ of the proton and neutron. Assuming isospin symmetry, $\Delta_u^{(n)}=\Delta_d^{(p)}$ and $\Delta_d^{(n)}=\Delta_u^{(p)}$,
\begin{align}
    \label{eq:up_down_axial_proton}
    a_0 &= \frac{1}{2}\left(\Delta_u^{(p)} + \Delta_d^{(p)} + \Delta_u^{(n)} + \Delta_d^{(n)}\right) = \Delta_u^{(p)} + \Delta_d^{(p)}\ ,
    \nonumber\\
    a_1 &= \frac{1}{2}\left(\Delta_u^{(p)} + \Delta_d^{(p)} - \Delta_u^{(n)} - \Delta_d^{(n)}\right) = 0\ .
\end{align}
We use $\Delta_u^{(p)}=0.777$ and $\Delta_d^{(p)}=-0.438$~\cite{DelNobile:2021wmp}, neglecting the strange-quark contribution. The isovector combination vanishes because the mediator couples universally to $u$ and $d$ quarks and isospin is conserved, so that the proton and neutron couplings are equal, independently of the numerical values of $\Delta_q^{(N)}$. Hence only $S_{00}^{\mathcal{L},\mathcal{T}}$ contributes.

Lastly, we also consider the sterile-neutrino dipole portal. This is a minimal up-scattering scenario in which the only BSM \enquote{ingredient} is a dipole operator coupling the active neutrino to a heavy sterile fermion $\chi$ and the photon~\cite{Marciano:1977wx,Lee:1977tib,Petcov:1976ff,Pal:1981rm}
\begin{equation}
    \label{eq:dipole_operator}
    \mathcal{L}_{\rm DP}=\frac{\mu_{\nu}}{2}\,\overline{\chi_R}\,\sigma_{\mu\nu}\,\nu_L\,F^{\mu\nu}
    + \text{H.c.}\ ,
\end{equation}
where $\mu_{\nu}$ is the transition dipole moment (with dimensions of inverse mass), $\sigma_{\mu\nu}=\frac{i}{2}[\gamma_\mu,\gamma_\nu]$, and $F^{\mu\nu}$ is the electromagnetic field-strength tensor. The subscripts $L$ and $R$ denote chirality, $\nu_L=P_L\nu$ being the left-handed active neutrino and $\chi_R=P_R\chi$ the right-handed component of the sterile fermion, with $P_{L,R}=(1\mp\gamma_5)/2$.
In full generality, the dipole coupling involves magnetic and electric dipoles and can depend on lepton flavor and on the sterile state. Here, we assume flavor universality and up-scattering to a single final-state fermion. The differential cross section in this case reads \cite{Shoemaker:2018vii}
\be
    \label{eq:dipole_xsec}
    \frac{\d \sigma^\text{DP}}{\d T_\text{nr}} = \frac{\pi \alpha_{{\rm EM}}^{2}}{m_{e}^{2}} Z^{2} F_\text{EM}^{2}(T_\text{nr})\left|\mu_{\nu}\right|^{2}\left[\frac{1}{T_\text{nr}}-\frac{1}{E_{\nu}} -\frac{m_{\chi}^{2}}{2 E_{\nu} T_\text{nr}m_{\mathcal{N}}} \left( 1- \frac{T_\text{nr}}{2E_{\nu}} +\frac{m_{\mathcal{N}}}{2E_{\nu}} \right) + \frac{m_{\chi}^{4}(T_\text{nr}-m_\mathcal{N})}{8E_{\nu}^{2}T_\text{nr}^{2} m_{\mathcal{N}}^{2}} \right]\, ,  
\ee
where $\alpha_\text{EM}=e^2/4\pi\simeq1/137$ is the fine-structure constant, $m_e$ the electron mass, $Z$ the atomic number of the target, $\mu_\nu$ is given in units of the Bohr magneton $\mu_B=e/2m_e$, and $F_\text{EM}$ is the electric-charge form factor, for which we use $F_{p,i}(\qtransfer^2)$ (the photon couples to the nuclear charge, so only the proton response enters).

The differential event rate for all the scenarios considered here follows from Eqs.~\eqref{eq:vector_xsec}, \eqref{eq:scalar_xsec}, \eqref{eq:axial_xsec} and \eqref{eq:dipole_xsec},
\be
\label{eq:DRS}
\frac{\d R}{\d T_{\rm nr}}
=
T_{\rm exp}\,N_T \, \epsilon(T_{\rm nr})\, 
\sum_i \eta_i
\int_{E_{\nu,i}^{\rm min}(T_{\rm nr})}^{E_\nu^{\rm max}}
\d E_\nu\,
\frac{\d \Phi}{\d E_\nu}\,
\frac{\d \sigma_i}{\d T_{\rm nr}}\,,
\ee
where the sum runs over the nine stable xenon isotopes with natural abundances $\eta_i$, $N_T$ is the total number of target nuclei, $T_{\rm exp}$ the exposure time, and $\epsilon(T_{\rm nr})$ is the experimental efficiency~\cite{LZ:2026axp}. For the atmospheric neutrino flux $\d\Phi/\d E_\nu$ we use Ref.~\cite{Vitagliano:2019yzm} (summed over flavors and $\nu$, $\bar\nu$, and normalized according to Ref.~\cite{Baxter:2021pqo}).
The lower intergation limit enforces the kinematic requirement that an incoming neutrino has sufficient energy to up-scatter to the heavy state and produce a given recoil energy \cite{Candela:2024ljb}
\be
\label{eq:Evmin_kinem}
E_{\nu}^{{\rm min}}(T_\text{nr}) = \frac{1}{2} \left(T_\text{nr} + \sqrt{2m_{\mathcal{N}}T_\text{nr}+(T_\text{nr})^{2}}\right)\left(1+ \frac{m_{\chi}^{2}}{2m_{\mathcal{N}}T_\text{nr}}\right).
\ee
The upper limit is the endpoint of the tabulated flux, $E_\nu^{\rm max} \simeq 2.3\times10^{5}$~GeV, and we have checked that for the LZ event best-fit masses the rate is saturated well below it, with more than 99\% of the events coming from $E_\nu\lesssim10$~TeV. 
\begin{figure*}
    \centering
    \includegraphics[width = 0.65 \linewidth]{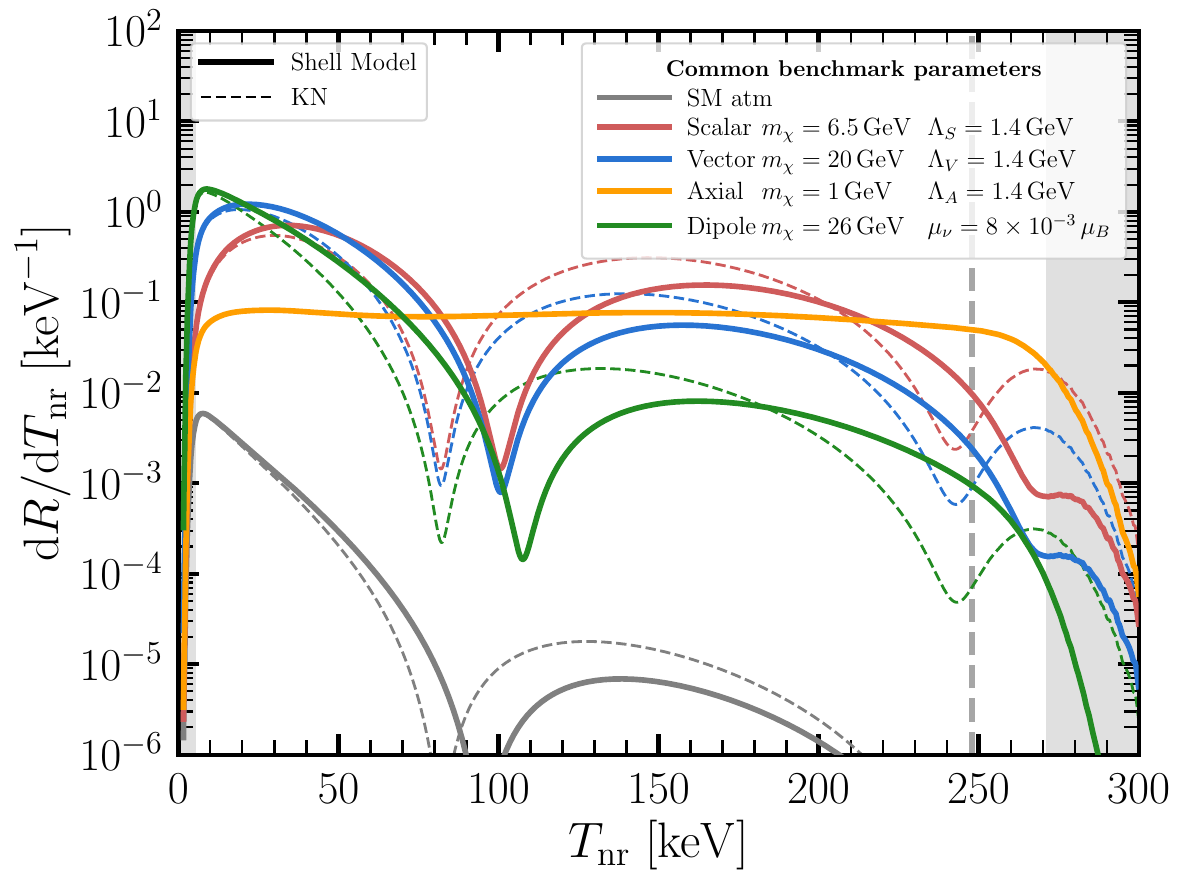}
   \caption{Expected differential event rate as a function of the nuclear recoil energy for the vector (blue), scalar (red), axial (orange) and sterile-dipole (green) scenarios, for an exposure of 2.84 tonne-years. The parameters $m_\chi$ and $\Lambda_X=m_X/g_X$ (or $\mu_\nu$) are fixed to the best-fit points of the one-event analysis, assuming a heavy mediator (see Sec.~\ref{sec:results}). Thick solid (thin dashed) lines show the results for the shell-model (KN) nuclear form factors. The gray lines show the rate expected from SM \cevns~events alone.}
    \label{fig:heavy}
\end{figure*}
It is worth noting that an upper cut on the neutrino energy below the end of the spectrum should not be imposed. It removes the high-energy tail and closes the kinematically available region for heavy states: a cut at $E_\nu^{\rm max}$ excludes all masses with $m_\chi\gtrsim\sqrt{2 \qtransfer E_\nu^{\rm max}}$, which is about 2~GeV for $E_\nu^{\rm max} = 10$~GeV and 22~GeV for $E_\nu^{\rm max} = 1$~TeV. It also artificially favors light $m_\chi$, which matters for the interpretation of the LZ event. For fixed couplings the rate falls steeply with $m_\chi$, because $E_\nu^{\rm min}\propto m_\chi^2$ moves into the steeply falling flux. Accounting for the full spectrum induce cross section suppressions that demand larger couplings for heavy states, in some cases close to the non-perturbative regime (see Sec.~\ref{sec:results}).  

Figure~\ref{fig:heavy} shows the expected differential event rates, as a function of the nuclear recoil energy, for the four scenarios considered here (vector, scalar, axial and dipole), and assuming the 2.84 tonne-year exposure. The vector (blue), scalar (red), axial (orange), and dipole (green) curves are obtained fixing $m_\chi$ and $\Lambda_X = m_X/g_X$ (or $\mu_\nu$) according to the best-fit points of the one-event analysis (assuming point-like interaction with a heavy mediator; see Sec.~\ref{sec:results} for more details). Thick solid and thin dashed lines correspond to the shell-model and KN nuclear form factors, respectively. The two choices differ most around the second diffraction minimum (Fig.~\ref{fig:form_factor}), which lies close to the energy of the LZ event. We also show as gray lines the expected rates assuming only Standard Model (SM) \cevns~events.

\begin{figure}
    \centering
\includegraphics[width=0.65\linewidth]{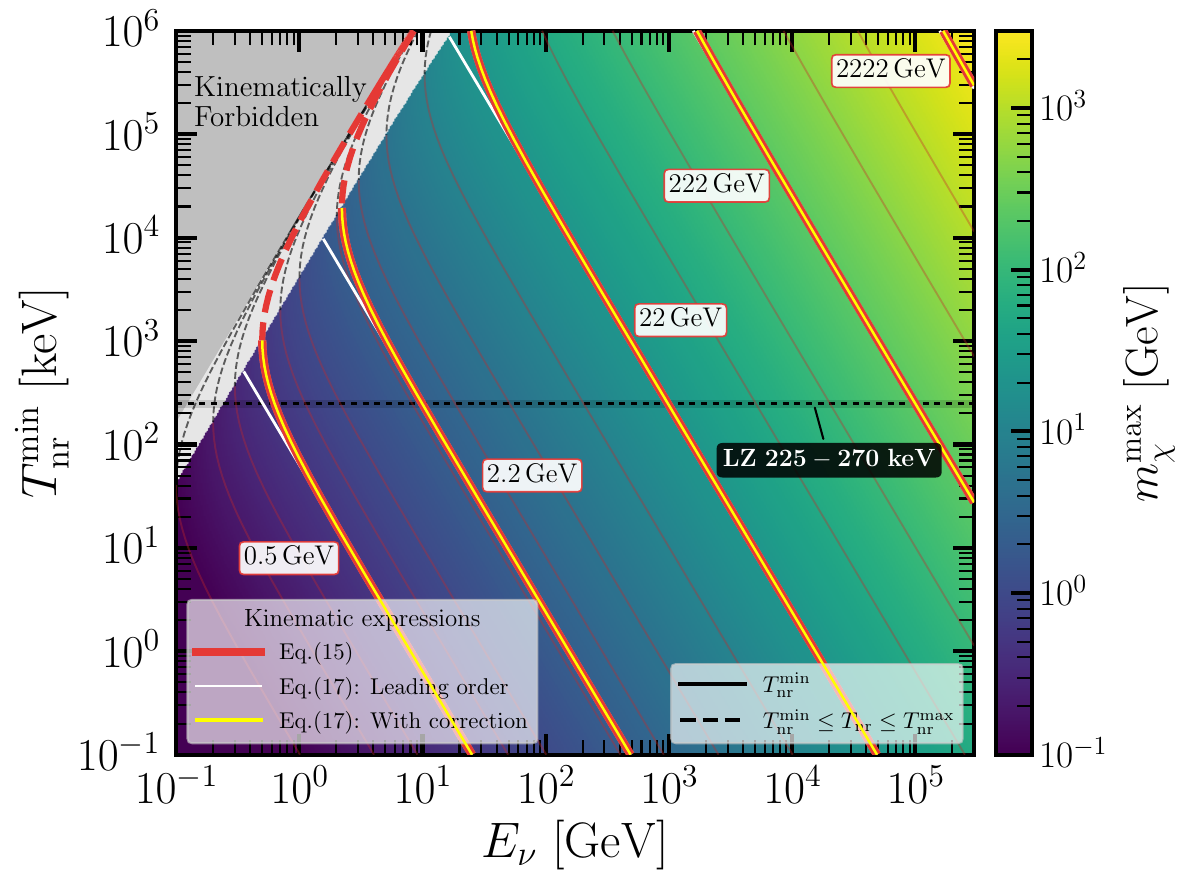}

    \caption{Minimal nuclear recoil energy, $T_{\rm nr}^\text{min}$, as a function of the incident neutrino energy $E_\nu$ for neutrino up-scattering on a xenon nucleus ($A=131$). The color shading represents the maximum kinematically accessible sterile-fermion mass, $m_{\chi,\max}$, restricted to the minimum-recoil region. Thick red contours show the exact kinematic results for $m_\chi=[0.5,2.2,22,222,2222]$ GeV, obtained by inverting Eq.~\eqref{eq:Evmin_kinem}.  White and yellow solid contours show the leading-order and corrected approximations of Eq.~\eqref{eq:Er_min_upscattering}, respectively, and are restricted to the minimum-recoil region. Solid contours correspond to the minimum recoil energy, $T_{\rm nr}^{\min}$, while dashed contours trace the higher-recoil boundary of the kinematically allowed interval, $T_{\rm nr}^{\min}\leq T_{\rm nr}\leq T_{\rm nr}^{\max}$. The light-gray region contains kinematically accessible recoil configurations that do not correspond to minimum-recoil solutions, whereas the dark-gray region is kinematically forbidden. The horizontal dashed line marks the LZ event at $T_{\rm nr}=248$ keV, while the shaded horizontal band indicates the $225$--$270$ keV recoil window.}
    \label{fig:kinematics}
\end{figure}

Before proceeding to the next section, we briefly discuss the kinematics of neutrino--nucleus up-scattering. This discussion will help clarify why, among the possible BSM neutrino scenarios, this mechanism appears potentially well suited to explaining the LZ result. Energy and momentum conservation imply
\begin{equation}
    \label{eq:recoil_energy_costheta}
    T_\text{nr}(\cos\theta)=\frac{2E_\chi^2-m_\chi^2-2E_\chi|\mathbf{p}_\chi|\cos\theta}{2(m_\mathcal{N}-E_\chi+|\mathbf{p}_\chi|\cos\theta)}\ ,
\end{equation}
where $\widehat{\mathbf{p}}_\nu\cdot\widehat{\mathbf{p}}_\chi=\cos\theta$. Here, $\theta$ is defined in the laboratory frame and $E_\chi=E_\nu-T_{\rm nr}$. The minimum recoil energy corresponds to forward emission of the outgoing fermion ($\cos\theta=1$). However, the maximum recoil energy does not necessarily correspond to backward emission ($\cos\theta=-1$), since the outgoing massive fermion may also propagate forward at the maximum-recoil endpoint, depending on the kinematics.

In the $m_\chi=0$ limit, we find $T_\text{nr}^\text{min}=0$ and $T_\text{nr}^\text{max}=2E_\nu^2/(m_\mathcal{N}+2E_\nu)$. Note that these are indeed the limits one finds in the conventional
$\nu+\mathcal{N}\to\nu+\mathcal{N}$
case. For up-scattering in the forward direction we instead find
\begin{equation}
    \label{eq:Er_min_upscattering}
    T_\text{nr}^\text{min}=\frac{(E_\chi-|\mathbf{p}_\chi|)^2}{2(m_\mathcal{N}-E_\chi+|\mathbf{p}_\chi|)}\simeq\frac{m_\chi^4}{8E_\nu^2 m_\mathcal{N}} \left( 1 + \frac{m_\chi^2(m_\mathcal{N}+E_\nu)}{2m_\mathcal{N} E_\nu^2}\right)\, ,
\end{equation}
where the small correction in the second term inside the parenthesis is needed when $m_\chi^2 \gtrsim 2m_\mathcal{N}E_\nu$.
Thus, in contrast to elastic forward scattering the minimum recoil energy in forward up-scattering does not vanish. This can be readily understood from the transferred momentum, $\boldsymbol{{\rm \mathbf{q}}}=\mathbf{p}_\nu-\mathbf{p}_\chi$.  For forward elastic scattering $\qtransfer^2=(E_\nu-E_\chi)^2=0$, i.e. $\mathbf{p}_\nu=E_\nu\widehat{z}=\mathbf{p}_\chi$. For forward up-scattering instead $|\mathbf{p}_\chi|\simeq E_\chi-\frac{m_\chi^2}{2E_\chi}\simeq E_\nu-\frac{m_\chi^2}{2E_\nu}$, where the latter approximation neglects the small nuclear recoil energy. This means that both $\mathbf{p}_\nu$ and $\mathbf{p}_\chi$ are still aligned but their magnitudes are no longer equal, resulting in a nonvanishing transferred momentum $\qtransfer\simeq m_\chi^2/(2E_\nu)$ at leading order.

Equation~\eqref{eq:Er_min_upscattering} is key to understanding why up-scattering is potentially well suited to explain the LZ one-event measurement. For a fixed neutrino energy, larger $m_\chi$ push the minimum recoil energy to larger values, so low-recoil bins are not populated, since kinematics enforces a nonzero momentum transfer even for forward scattering. Conversely, recoils of $T_\text{nr}=250$~keV are kinematically accessible only for $m_\chi\lesssim(8m_\mathcal{N}T_\text{nr}E_\nu^2)^{1/4} \simeq 0.7\sqrt{E_\nu/\text{GeV}}$~GeV. For $E_\nu = [10,10^3,10^5,10^7]\,\text{GeV}$ we find $m_\chi = [2.2,22,222,2222]\,\text{GeV}$. Figure~\ref{fig:kinematics} illustrates this behavior. For a given neutrino energy, the heavier the sterile fermion, the higher the minimum recoil energy, so that the region of low recoil energies is kinematically inaccessible. Conversely, producing a recoil of $\sim250$~keV with $m_\chi\gtrsim1$~GeV requires neutrino energies in the GeV--TeV range, and heavier fermions need correspondingly larger energies. Thus, if one seeks to explain the LZ event while kinematically forbidding lower energy events, one would expect sterile fermion masses $\mathcal{O}\sim 10^{1}-10^{2}$~GeV to be preferred if neutrinos up to $E_\nu^{\rm max} \simeq 10^{5}$~GeV are allowed. We will see below, that our analysis of the LZ event indeed prefers heavy sterile masses, originating from very high energy atmospheric neutrinos, but that the cross section requirements push the coupling into non-perturbative regime.

\section{Results}
\label{sec:results}
The LZ dataset~\cite{LZ:2026axp} covers the nuclear recoil energy range $[5.4,270]\;\text{keV}_\text{nr}$. The results are presented in terms of position-corrected prompt and delayed scintillation signals, S1c and S2c, measured in photons detected (phd).  For our purposes, however, working with nuclear recoil energy $T_\text{nr}$ is more convenient, thus we adopt the following strategy. Using the $1\sigma$ statistical uncertainty of the energy of the observed event, $23\;\text{keV}_\text{nr}$, we split the range into two bins of unequal width: a first bin restricted to $3<\mathrm{S1c}<500~\mathrm{phd}$, $2.75<\log_{10}(\mathrm{S2c}/\mathrm{phd})<4.15$, and $\mathrm{S2}>645~\mathrm{phd}$, which is approximately equivalent to $\mathcal{B}_1=[5.4,225]$~keV, consistent with the background-only hypothesis, and a second bin, $500<\mathrm{S1c}<600~\mathrm{phd}$, corresponding to $\mathcal{B}_2=[225,270]$~keV, in which one event above the expected background is observed~\cite{LZ:2026axp}. 
This is of course a \textit{working choice} since one could adopt a different strategy. For instance, one could instead add the statistical and systematic energy uncertainties in quadrature (assuming they are uncorrelated), which narrows $\mathcal{B}_1$ (edge at $248-32.5 = 215.5$~keV) and widens $\mathcal{B}_2$. 
We stress that although results are expected to have a dependence on this procedure, differences between them (a few percent in terms of number of events per bin) will not change the overall conclusions. 

Assuming the background-only hypothesis ($H_0$) and the up-scattering signal hypothesis ($H_1$), in principle, the full dataset can be used in two complementary ways. (i), as a conservative test of whether $H_1$ is compatible with the data: both bins are used to exclude regions of parameter space of the scenarios discussed in Sec. \ref{sec:theoretical_frame}. Here the observation in $\mathcal{B}_1$ dominates, as the expected number of BSM events in $\mathcal{B}_2$ is smaller than in $\mathcal{B}_1$ for the couplings of interest.  
(ii), we use $\mathcal{B}_2$, while respecting the constraint from $\mathcal{B}_1$, to ask whether up-scattering can account for the single event, and we infer the favored regions of parameter space.

To derive limits and favored regions we use a binned Poisson likelihood analysis with the test statistic 
\begin{equation}
    \chi^2(\boldsymbol{\theta}) = 2\sum_{i=1}^{2}\left[R_i^{\rm th}(\boldsymbol{\theta})-R_i^{\rm obs}+R_i^{\rm obs}\ln\frac{R_i^{\rm obs}}{R_i^{\rm th}(\boldsymbol{\theta})}\right]\ ,
    \label{eq:chi2_LZ}
\end{equation}
where $\boldsymbol{\theta}$ denotes the relevant set of BSM
parameters, and the predicted number of events is
\begin{equation}
    R_i^{\rm th}(\boldsymbol{\theta}) = B_i + R_i^{\rm BSM}(\boldsymbol{\theta})\ , \qquad B_i = B_i^{{\rm non}\text{-}\nu}+B_i^{{\rm SM}\text{-}\nu}\ .
    \label{eq:Nth_LZ}
\end{equation}
Here, $B_i$ denotes the total expected experimental
background, including the SM atmospheric \cevns~contribution ($B_i^{{\rm SM}\text{-}\nu}$), while $R_i^{\rm BSM}(\boldsymbol{\theta})$ is obtained by integrating Eq.~\eqref{eq:DRS} over the recoil energy range for each bin ($i = \mathcal{B}_1, \mathcal{B}_2$).
The observed counts and expected backgrounds are
\begin{equation}
    R_1^{\rm obs}=1709,\ B_1=1712\ ;\qquad R_2^{\rm obs}=1,\ B_2=0.0106\ ,
    \label{eq:LZ_bkg_bins}
\end{equation}
as inferred from Tab. I and Fig. 5 of Ref.~\cite{LZ:2026axp}.
 We adopt the fixed total background expectations $B_1 = 1712$ and $B_2 = 0.0106$, which include the
SM neutrino contributions.
Assuming SM interactions only, in $\mathcal{B}_1$ we find 0.11 (0.09) atmospheric \cevns~events incorporating shell model (KN) nuclear form factors, in excellent agreement with the LZ Collaboration. Instead, in $\mathcal{B}_2$ we  estimate $\sim 10^{-6}$ events, aligned with the negligible SM \cevns~backgrounds reported by LZ. This serves as a validation test of our present calculations.
\begin{figure*}[t]
    \centering
    \includegraphics[scale=0.44]{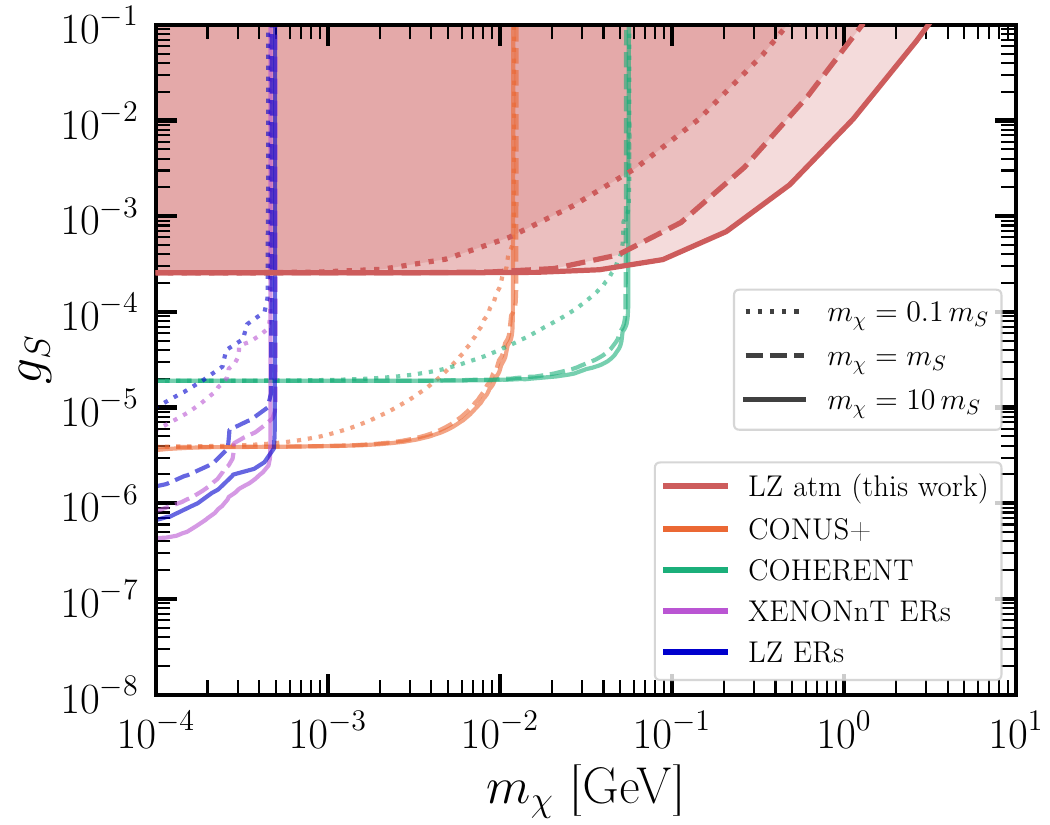} \includegraphics[scale=0.44]{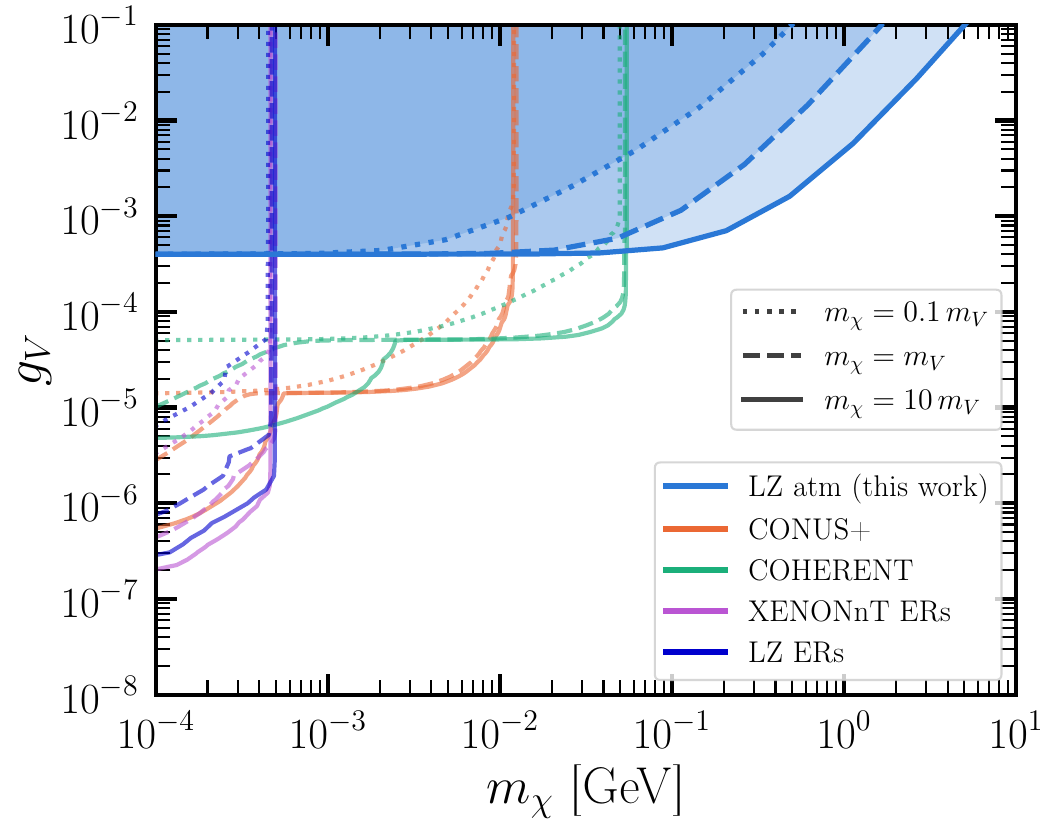} \\
    \includegraphics[scale=0.44]{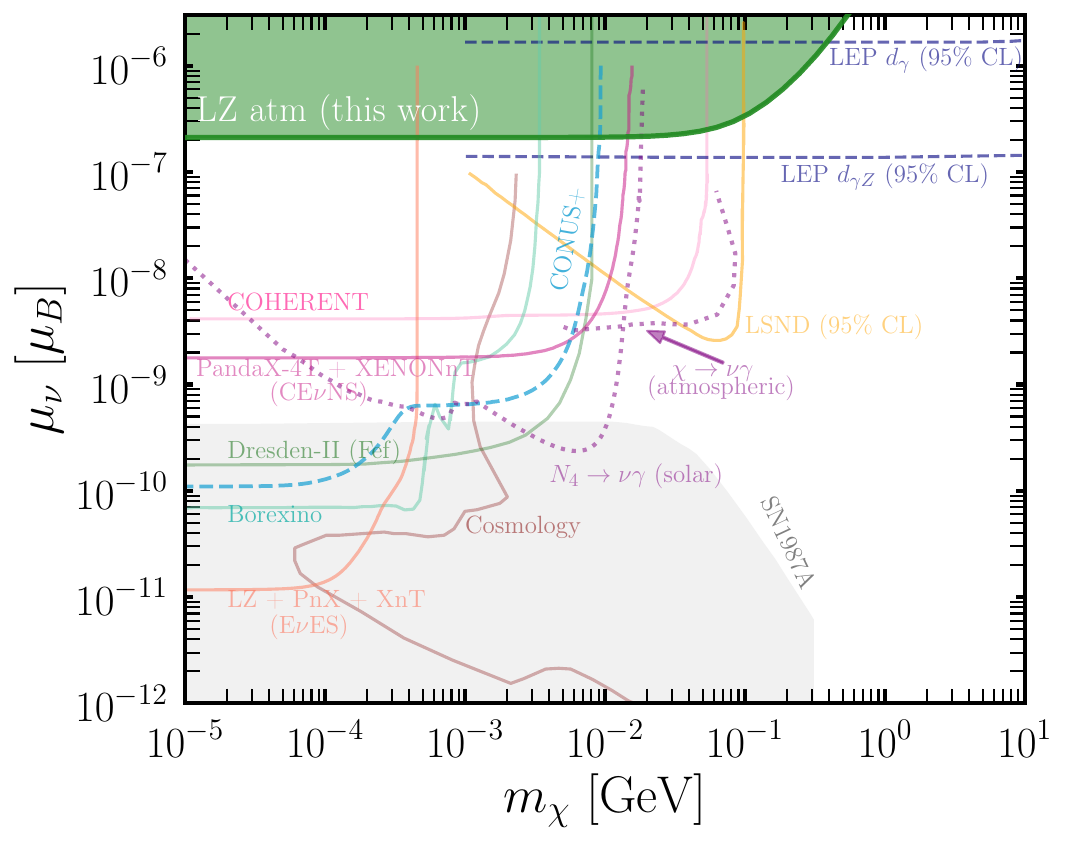}
    \includegraphics[scale=0.44]{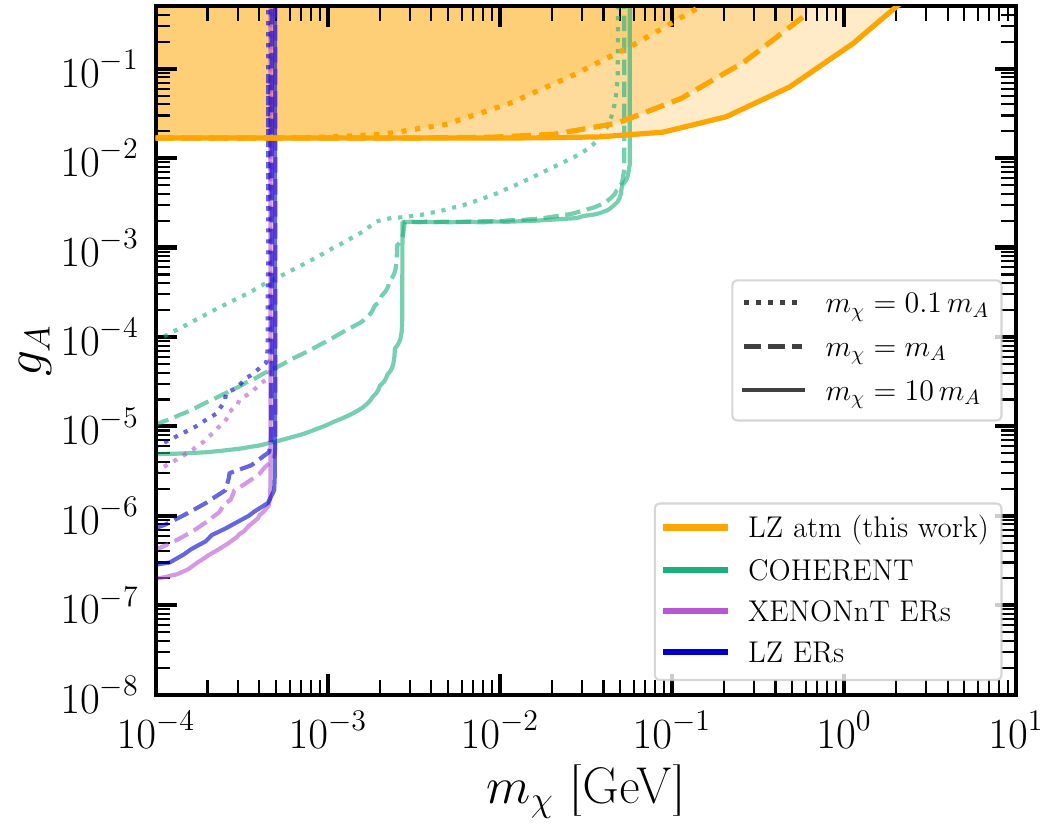} 
    \caption{Upper limits at 90\% CL ($\Delta\chi^2=4.61$) from the LZ data~\cite{LZ:2026axp}: on the scalar and vector couplings (top row), on the sterile-dipole moment $\mu_\nu$ (bottom left) and on the axial coupling (bottom right), as a function of the sterile-state mass $m_\chi$. For the scalar, vector and axial cases, the final-state mass is fixed to $m_\chi=0.1\,m_X$ (dotted), $m_X$ (dashed) and $10\,m_X$ (solid). Results are displayed along with constraints from dedicated \cevns~measurements at COHERENT~\cite{Candela:2023rvt,Miranda:2021kre,DeRomeri:2022twg} and CONUS+~\cite{DeRomeri:2025csu} as well as with LZ and XENONnT constraints from electron-recoil data~\cite{Candela:2024ljb,A:2022acy,Giunti:2023yha,DeRomeri:2024hvc}. In the dipole case further constraints are included (see text for more details).}
    \label{fig:bounds}
\end{figure*}

\begin{figure*}[t]
    \centering
    \includegraphics[scale=0.44]{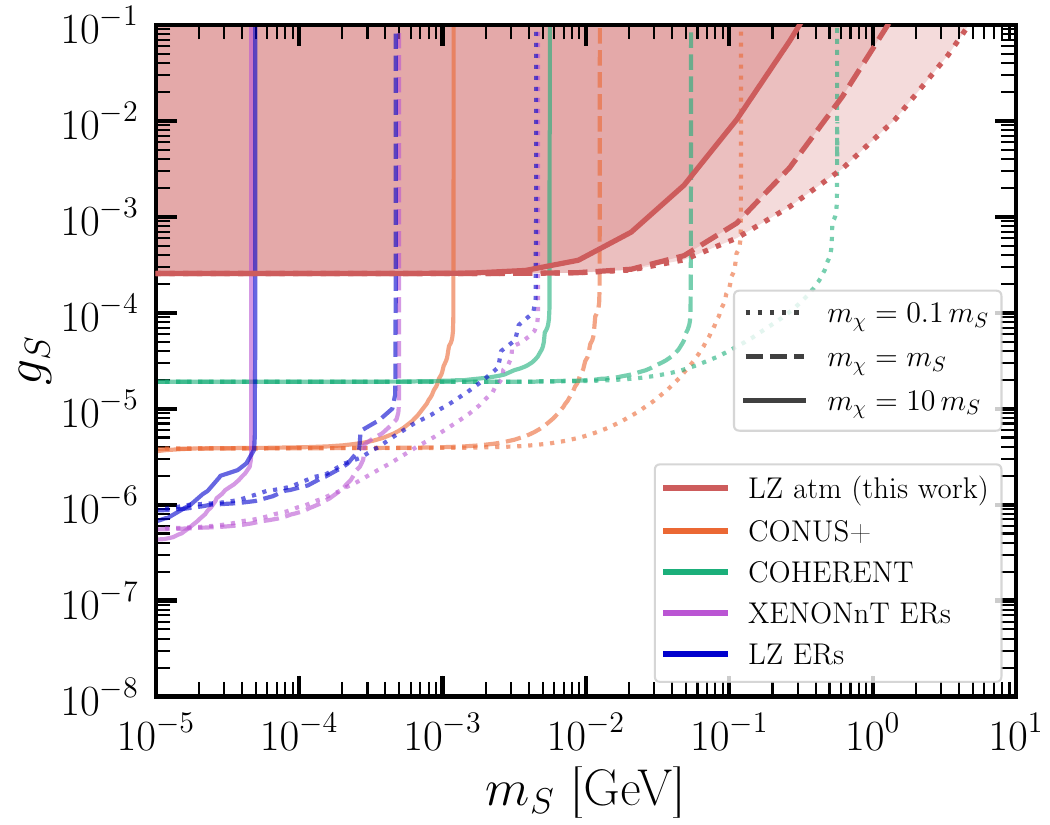} \includegraphics[scale=0.44]{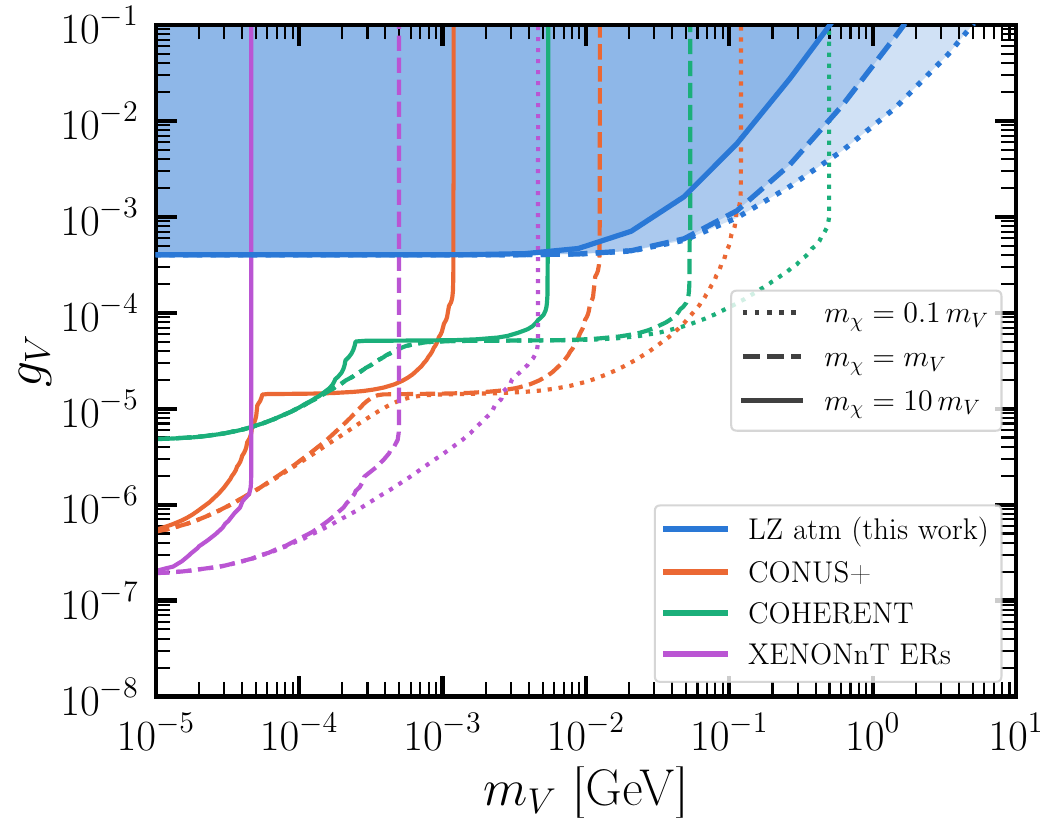} \\
    \includegraphics[scale=0.44]{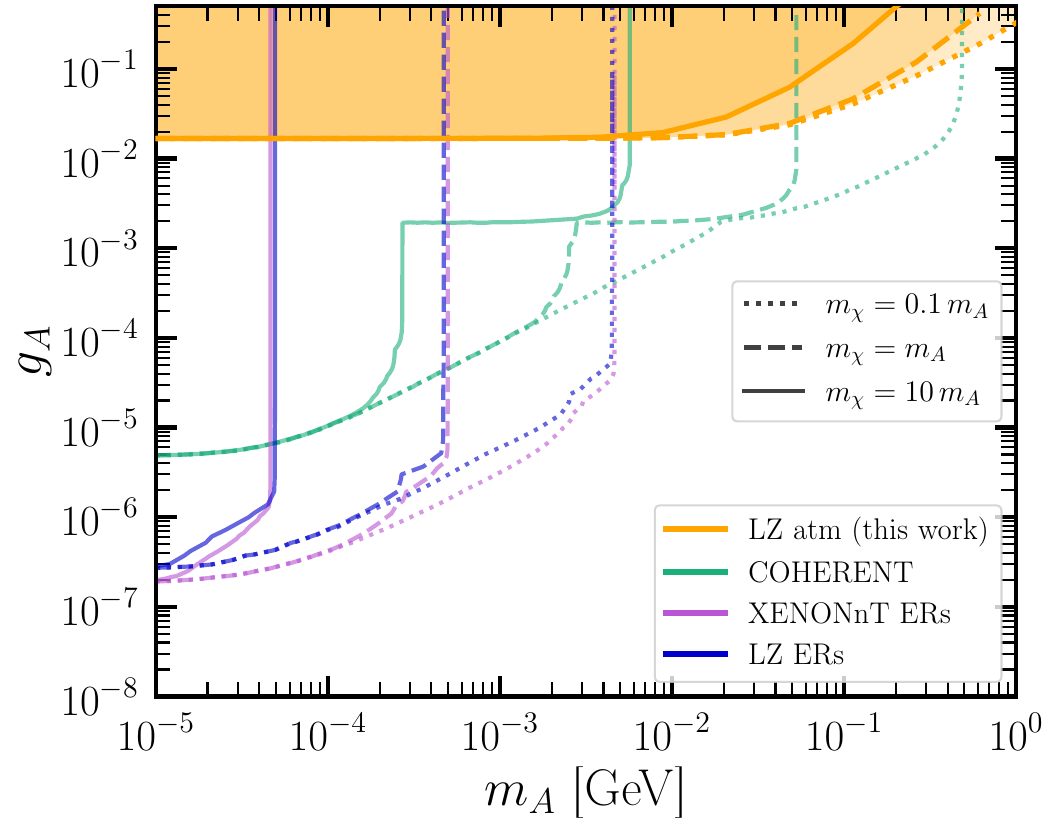} 
    \caption{Same as Fig.~\ref{fig:bounds}, but as a function of the mediator mass $m_X$: 90$\%$ CL ($\Delta \chi^2 = 4.61$) bounds based on the LZ data~\cite{LZ:2026axp} for scalar (top left), vector (top right) and axial (bottom) mediated scattering. In all scenarios, the final-state mass is fixed according to: $m_{\chi} = 0.1\,m_{S}$ (dotted), $m_{\chi} = m_{S}$ (dashed), and $m_{\chi} = 10\,m_{S}$ (solid).}
    \label{fig:bounds_mX}
\end{figure*}

Following strategy (i), we determine the regions of parameter space that LZ data rule-out for all the scenarios discussed in Sec. \ref{sec:theoretical_frame}. Our main results are shown in Figs.~\ref{fig:bounds} and~\ref{fig:bounds_mX}. In Fig.~\ref{fig:bounds} we show bounds in the sterile-state mass--coupling plane, with  the mediator mass correlated with the final-state mass as indicated in the legends. At low mediator masses, LZ results are not competitive with \cevns-dedicated experiments nor with DM detectors measurements in the electron channel, including LZ itself. This is somewhat expected: COHERENT and CONUS+ probe much smaller momentum transfers, thus implying larger cross sections in those regions of parameter space. Electron-channel measurements at XENONnT and LZ profit from much larger solar neutrino fluxes, $pp$ and $^7\text{Be}(0.8613\;\text{MeV})$. At higher masses these experiments lose sensitivity because of kinematic thresholds, and it is in this region that the LZ bounds, that are dominated by $\mathcal{B}_1$, become relevant. In particular, for scalar and vector interactions LZ data already rule out previously allowed regions with $m_\chi \gtrsim 60$~MeV and, interestingly, are comparable to or even slightly extending the projected sensitivity reach at DUNE~\cite{Candela:2024ljb}.

The sterile dipole case is an exception. We confront this scenario with a broad set of existing bounds. Constraints from nuclear recoils come from COHERENT CsI and LAr~\cite{Miranda:2021kre,DeRomeri:2022twg}, from Dresden-II~\cite{AristizabalSierra:2022axl}, for which we adopt the iron-filter quenching factor model, and from the combined \cevns~analysis of XENONnT and PandaX-4T data~\cite{DeRomeri:2024hvc}. Electron-recoil measurements enter through the combined analysis of XENONnT, LZ and PandaX-4T data~\cite{A:2022acy,Giunti:2023yha,DeRomeri:2024hvc} and through Borexino~\cite{Brdar:2020quo,Plestid:2020vqf}. We also include the bounds from LSND and LEP~\cite{Magill:2018jla}, and those on the radiative decay $\chi\to\nu\gamma$ derived from solar neutrinos (Borexino and Super-Kamiokande)~\cite{Plestid:2020vqf,Plestidlumsolnu} and from atmospheric neutrinos (Super-Kamiokande)~\cite{Gustafson:2022rsz}. Finally, for comparison, we display the limits from SN1987A~\cite{Magill:2018jla,Chauhan:2024nfa} and from cosmology, namely BBN~\cite{Magill:2018jla,Brdar:2020quo} and CMB measurements of $\Delta N_\mathrm{eff}$~\cite{Brdar:2020quo}. 
For the masses at which the LZ bound overtakes those of COHERENT~\cite{Miranda:2021kre,DeRomeri:2022twg}, CONUS+~\cite{DeRomeri:2025csu} and other low-energy probes, the strongest constraints come from LSND and from LEP~\cite{Magill:2018jla}. In the latter, the heavy state is produced in $e^+e^-\to\nu\bar\chi$ followed by $\chi\to\gamma\nu$, giving a single-photon plus missing-energy signature. This process can be mediated by either a photon or a $Z$ boson, so two LEP bounds exist; since nuclear scattering probes only the photon dipole, if we consider only photon exchange the LZ bound is stronger than both LSND and LEP for $m_\chi \gtrsim 100$~MeV. This comparison, however, has to be done with some care. The limits apply to an effective coupling $\mu_\nu$, which at the cross-section level is in general a complicated function of the underlying couplings, the baseline and the neutrino mixing matrix elements, and, for atmospheric neutrinos, may include matter effects (for earlier work on the subject see e.g. Ref. \cite{Grimus:1997aa}, for more recent work see Refs. \cite{Miranda:2021kre,AristizabalSierra:2021fuc,Ternes:2025lqh}). In other words, constraints apply over an \textit{effective} coupling whose structure depends upon the experimental environment. We should however note that expected differences are small and likely not visible on the large parameter space here explored, spanning many orders of magnitude in a log-log plane. We do not attempt to compare the bounds at the level of fundamental couplings, and simply conclude that in terms of the effective coupling LZ data are competitive with the photon-only LEP bound for $m_\chi\gtrsim100$ MeV.

Figure~\ref{fig:bounds_mX} shows the same bounds in the mediator mass-coupling plane, with the final-state mass fixed relative to the mediator mass, $m_\chi=0.1\,m_X$, $m_X$ and $10\,m_X$. At low mediator masses, the constraints are dominated by COHERENT, CONUS+ and the electron-channel measurements. As these lose sensitivity toward larger masses, LZ takes over and rules out previously allowed regions of parameter space, probing $m_X\gtrsim 6$~MeV for $m_\chi=10\,m_X$ and $m_X\gtrsim 500$~MeV for $m_\chi=0.1\,m_X$. Dark matter detectors in this energy window are therefore already sensitive to regions of the up-scattering parameter space that have not been tested before. These bounds are solid: they rely only on the absence of an excess in $\mathcal{B}_1$, and not on the interpretation of the 248~keV event. They would therefore remain valid, essentially unchanged, even if the event were to disappear. With larger exposures and a better understanding of the experimental effects, LZ and other DM detectors have the potential to provide world-class limits in this intermediate energy regime. 
Let us also add that $^8$B solar neutrinos, for which \cevns~has now been observed in xenon detectors (LZ~\cite{LZ:2025igz} itself, XENONnT~\cite{XENON:2024ijk,XENONCollaboration:2026xfz} and PandaX-4T~\cite{PandaX:2024muv}), can also probe these scenarios, but only for lighter final-state fermions: their energy of at most 16~MeV limits the threshold to $m_\chi\lesssim\mathcal{O}(10)$~MeV for the few-keV recoils they induce. The high-recoil-energy data analyzed here extend the reach to $m_\chi\gtrsim$~GeV through the atmospheric flux, so the two data sets are complementary.

\begin{figure*}[t]
    \centering 
    \includegraphics[width=\textwidth]{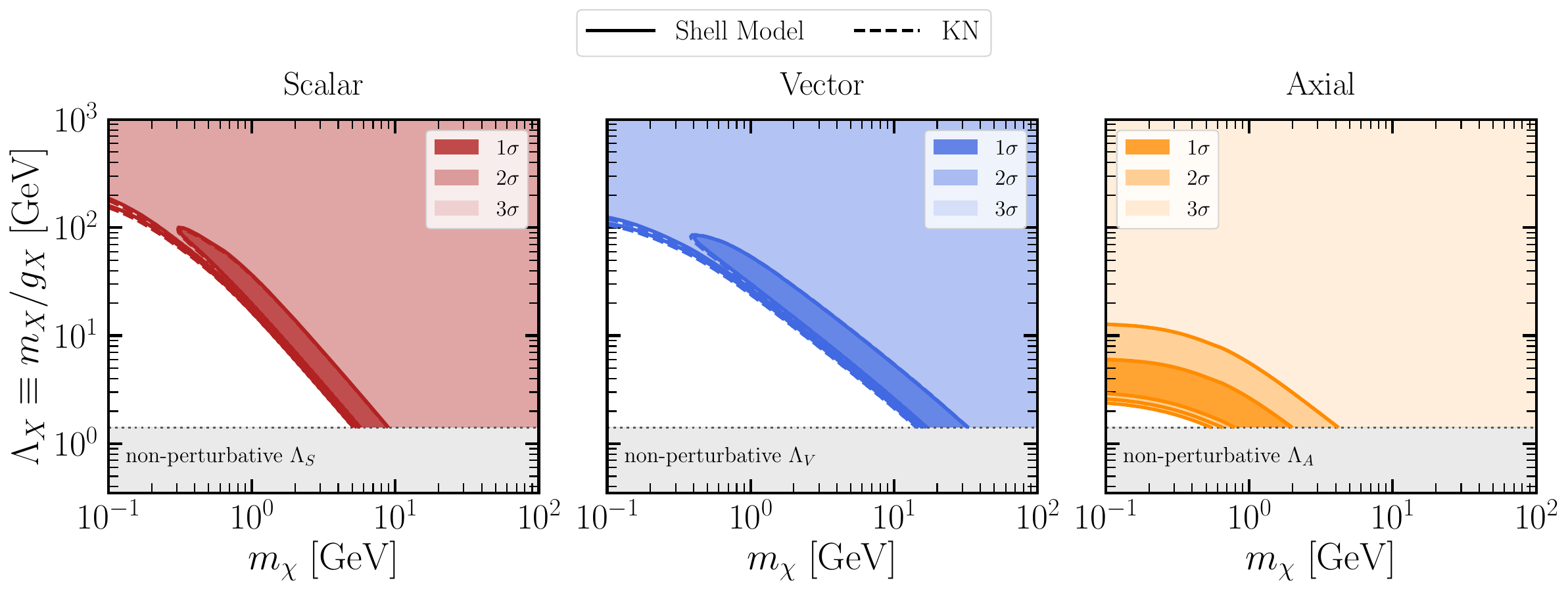}
    \caption{Regions of the $m_\chi$--$\Lambda_X$ plane ($\Lambda_X=m_X/g_X$) preferred by the one-event analysis for heavy-mediator up-scattering, for the scalar (left), vector (center) and axial (right) interactions. The shaded bands, from darkest to lightest, show the $1\sigma$, $2\sigma$ and $3\sigma$ regions ($\Delta\chi^2\le2.30,\ 6.18$ and $11.83$) with respect to the best-fit point. Solid (dashed) contours correspond to the shell-model (KN) form factors. The gray regions below the horizontal cuts are excluded by perturbativity (for details, see the text).}
    \label{fig:EFT}
\end{figure*}
\begin{table*}[h!]
\centering
\normalsize
\renewcommand{\arraystretch}{1.3}
\setlength{\tabcolsep}{8pt}
\begin{tabular}{@{}llcccccc@{}}
\toprule
\toprule
\multirow{2}{*}{Interaction}
& \multirow{2}{*}{form factor}
& \multirow{2}{*}{$m_\chi$ [GeV]}
& \multirow{2}{*}{$\chi^2_{\min}$}
& \multicolumn{2}{c}{Total events}
& \multicolumn{2}{c}{BSM events} \\
\cmidrule(lr){5-6}\cmidrule(lr){7-8}
& & & & $R_{\rm I}$ & $R_{\rm II}$
& $R_{\rm I}^{\rm BSM}$ & $R_{\rm II}^{\rm BSM}$ \\
\midrule
\multirow{2}{*}{Scalar}
& KN          & 6.85 & 1.66 & 1744 & 0.31 & 32.32 & 0.30 \\
& shell model & 6.93 & 1.17 & 1742 & 0.43 & 29.54 & 0.42 \\
\midrule
\multirow{2}{*}{Vector}
& KN          & 21.15 & 3.85 & 1747 & 0.09 & 35.13 & 0.08 \\
& shell model & 21.83 & 3.37 & 1747 & 0.12 & 35.20 & 0.11 \\
\midrule
Axial-vector & shell model & 1.25 & 0.06 & 1719 & 0.96 & 6.68 & 0.95 \\
\bottomrule
\bottomrule
\end{tabular}
\caption{Best-fit parameters and predicted event counts in the two bins $\mathcal{B}_1$ and $\mathcal{B}_2$ for the scalar, vector and axial-vector interactions, using the KN form factor and the shell-model nuclear responses. In all cases the best fit lies at the perturbativity boundary, $\Lambda_X \simeq1.41$~GeV. The total events include the background, the SM contribution and the BSM contribution; the last two columns contain only the BSM contribution.}

\label{tab:bestfit_EFT}
\end{table*}

As anticipated, we have also tested these hypotheses against the one-event observation. Adopting strategy (ii), we determine the regions of parameter space that are potentially consistent with the data in a two-parameter fit to $\Lambda_X^{-1}\equiv g_X/m_X$ and $m_\chi$, working in the effective limit $m_X^2\gg\qtransfer^2$, in which the rates depend on $m_X$ and $g_X$ only through their ratio. This is the most favorable choice for the one-event interpretation: for lighter mediators the propagator $1/(m_X^2+\qtransfer^2)^2$ is largest at small momentum transfer, so the rate would increase even more at low recoil energies, which is the opposite of what is needed to explain an event at $\sim 248$~keV without conflicting with the observation in $\mathcal{B}_1$. Perturbativity is ensured by requiring $g_X\le\sqrt{4\pi}$ and $m_X\ge5$~GeV, the latter being much larger than the maximum momentum transfer, $\qtransfer_{\rm max}\simeq257$~MeV; this implies $\Lambda_X\ge5/\sqrt{4\pi}\simeq 1.41$~GeV. The $1\sigma$--$3\sigma$ regions ($\Delta\chi^2=2.30,\,6.18,\,11.83$) in the $m_\chi$--$\Lambda_X$ plane are shown in Fig.~\ref{fig:EFT} for the scalar, vector and axial cases. No good fit is found for the dipole, whose cross section, dominated by the $1/T_{\rm nr}$ term, would predict far more events at low recoil energies than observed in $\mathcal{B}_1$.

As discussed in Sec.~\ref{sec:theoretical_frame}, the second minimum of the KN form factor falls within the region of interest and can thus artificially suppress the event rate.
To quantify this, we have performed the analysis using both the KN and the shell-model form factors. Despite this behavior, as shown in Fig.~\ref{fig:EFT} and Table~\ref{tab:bestfit_EFT}, the results differ only slightly. For the BSM prediction, differences on the event rate in the first bin are small, with the larger difference being of the order of 10\% in the scalar case. For the vector case those differences fall below the per mille level. In the second bin those differences grow larger in both the scalar and vector cases, about $\sim 20\%$. The axial case does not depend on this choice, since it only involves the shell-model spin structure functions. All in all, nuclear effects do have an impact on rates, but not at the level required to alter the conclusions of the statistical analysis.

As shown in Fig.~\ref{fig:EFT}, increasing the final-state fermion mass requires larger couplings \(g_X\), or equivalently smaller values of \(\Lambda_X\). This behavior is expected: the neutrino energy threshold grows as $m_\chi^2$, so heavier states can only be produced by the steeply falling high-energy tail of the atmospheric flux, and the one-event constraint then demands stronger interactions to compensate. 
As anticipated in Sec. \ref{sec:theoretical_frame}, moving toward heavier final-state fermion masses reduces the available phase space and so the cross section. With the one-event constraint, $g_X$ has to increase to compensate for the shrinking of parameter space. Indeed, the best-fit values of $\Lambda_X$ lie at the boundary of the perturbative region, imposed in the scan. Without this requirement, the fit would prefer even smaller $\Lambda_X$, deep inside the non-perturbative regime. In the end, although all three scenarios can describe the two-bin event counts, their best-fit points lie at the boundary of the perturbatively allowed parameter space. At higher confidence levels, the allowed regions extend to larger $\Lambda_X$, as expected for a single-event measurement.
\section{Conclusions}
\label{sec:conclusions}
In this work, we have examined the recent LZ dataset focusing on an extended nuclear-recoil window, up to approximately 270~keV, through the lens of neutrino up-scattering to a heavy sterile fermion $\chi$ induced by atmospheric neutrinos. We have explored the potential of the full LZ dataset in constraining vector, axial-vector, scalar, and sterile dipole interactions. To this aim, we  divided the data into two bins. The first bin is approximately equivalent to $5.4\leq T_{\rm nr}<225~\mathrm{keV}$ and involves only well-understood backgrounds and detector response. The second bin corresponds to $225 \leq T_{\rm nr} \leq 270~{\rm keV}$ and contains the single anomalous event reported by LZ, with characteristics consistent with a nuclear recoil of about 248 keV.

Our main result is that the absence of a signal in the first bin provides powerful constraints on scenarios with new mediators, in regions of parameter space not explored before. This is particularly the case for final-state masses $\gtrsim1$~GeV, which lie beyond the kinematic reach of reactor, spallation-source and solar neutrinos. These bounds depend only on the first bin, so they remain valid independently of the nature of the 248~keV event, even if it were to disappear. They are also complementary to those that can be obtained from the $^8$B solar-neutrino nuclear-recoil signal in xenon detectors, with the atmospheric flux probing new sterile fermions at the GeV scale and above.
For the dipole portal, the bounds are placed on the effective dipole moment $\mu_\nu$, which is in general a complicated function of baseline, mixing-matrix elements (including matter effects) and possible CP-violating phases. In terms of $\mu_\nu$, LEP gives the strongest constraint when both photon and $Z$ exchange are included, whereas if only photon exchange is considered, the LZ bound here derived is stronger for $m_\chi \gtrsim 100$~MeV.

Using data from both bins, we have then investigated whether up-scattering can account for the observed event. Reconciling a recoil at 250~keV with the absence of events at lower energies requires a spectrum shifted to high recoil energies, which kinematics provides only for heavy final states, since the minimum recoil energy grows as $m_\chi^4$. Heavy sterile states, in turn, are suppressed by the steeply falling atmospheric flux at the required neutrino energies, so that very large couplings are needed. For the scalar, vector and axial mediators the best-fit points, with $m_\chi\simeq1$-$20$~GeV, artificially lie at the edge of the perturbative region, whereas the dipole does not provide a good fit. Within the simplified models considered, up-scattering, while possible, is not a compelling explanation of the event. More generally, it seems that a neutrino-related explanation of the LZ one-event observation is unlikely.

This conclusion should be qualified. The momentum transfer associated with the event, $\qtransfer \approx0.25$~GeV, is large, and the coherent rate lies close to a zero of the nuclear form factor, which suppresses it and makes it sensitive to the form-factor model. Incoherent scattering off individual nucleons is not suppressed in this way and can contribute significantly to the rate at these recoil energies, which could, in principle, relax the couplings required to explain the event. The present work is thus a first step toward a more complete study of neutrino up-scattering (and in general, atmospheric neutrino-nuclear scattering) at LZ and other direct detection experiments.

In summary, the LZ data in this energy window provide useful information to constrain neutrino new physics, filling the gap between the low-energy regime, dominated by \cevns-related probes, and the high-energy regime, dominated by neutrino beams from pion decay in flight and by collider searches. More importantly, our results show that well-understood background measurements in this regime are a valuable tool to constrain BSM scenarios, and that their sensitivity will grow with exposure, as the expected SM atmospheric \cevns~yield, about $0.1$ events in the first bin, indicates that the atmospheric neutrino fog is not far away. A dedicated analysis by the experimental collaborations, using the full S1c--S2c information, would further sharpen these bounds. 
\section{Acknowledgments}
The work of D.A.S. is supported by ANID grant \enquote{Fondecyt Regular} 1260595. VDR~acknowledges financial support from CIDEXG/2022/20 (Generalitat Valenciana), and the Spanish grants PID2023-147306NB-I00 and CEX2023-001292-S (MCIU/AEI/10.13039/501100011033). IMS is supported
by the U.S. Department of Energy under the award number DE-SC0020262.  D.K.P. acknowledges support from the European Union’s Horizon Europe research and innovation programme under the Marie Skłodowska‑Curie Actions grant agreement No.~101198541 (neutrinoSPHERE). D.K.P. benefits from the scientific environment of the Cluster of Excellence "Quantum Universe" funded by the Deutsche Forschungsgemeinschaft (DFG, German Research Foundation) under Germany’s Excellence Strategy (EXC 2121 “Quantum Universe”-390833306).

\bibliographystyle{utphys}
\bibliography{biblio}
\end{document}